\documentstyle[11pt]{article}

\begin{document}

\baselineskip25pt
\centerline{\Large\bf Analog to Digital Conversion in Physical Measurements}

\vspace{15mm}
\centerline{\large T. Kapitaniak$^{(1,2)}$, K. \.Zyczkowski$^{(1,3)}$, 
U. Feudel$^{(1,4)}$ and C. Grebogi$^{(1)}$}
\centerline{$^{(1)} $Institute for Plasma Research, University of Maryland,
College Park, MD 20742, USA}
\centerline{$^{(2)}$ Katedra Dynamiki, Politechnika {\L}{\'o}dzka, 
ul. Stefanowskiego 1/15, 90-924 {\L}{\'o}d{\'z}, Poland }
\centerline{$^{(3)}$ Instytut Fizyki im. Smoluchowskiego,
  Uniwersytet Jagiello{\'n}ski, \\
 ul. Reymonta 4, 30-059 Krak{\'o}w, Poland}
\centerline{$^{(4)}$ Institut f\"ur Physik, Universit\"at Potsdam, 
PF 601553, 14415 Potsdam, Germany}

\vspace{10mm}

{\it
There exist measuring devices where an analog input is converted into a 
digital output. Such converters can have a nonlinear internal dynamics. 
We show how measurements with such converting devices can 
be understood using concepts from symbolic dynamics. 
Our approach is based on a nonlinear one-to-one mapping between 
the analog input and the 
digital output of the device. We analyze 
the Bernoulli shift and the tent map which are realized in specific 
analog/digital converters. Furthermore, we discuss
the sources of errors that are inevitable in physical realizations of 
such systems and suggest methods for error reduction.
 }

\vspace{10mm}

$PACS$ $no. : 05.45.+b$.

\vspace{1cm}

Progress in physics is based on the intimate relationship between
experiment and theory. Experiments involve measurements
in which the precision of the measured quantities depends essentially
on the particular characteristics of the measuring devices. Often this 
process of measurement involves a conversion of the analog measured 
quantity into a digital number by means of a converter. 
For such analog/digital converters, there exists an internal 
relationship between the quantity to be measured (analog input),
and the result of the measurement process (digital output). We assume 
that the conversion of the analog signal into a digital number is the 
sole process in such converters and we neglect  
errors associated with the input analog signal itself. Thus, 
the dynamics of the input-output
relationship determines the whole conversion process and it is responsible for 
the precision of the conversion itself. This problem
is especially relevant for converters based on nonlinear processes,
for which it is known that the errors grow exponentially fast.
In this paper, we analyze the properties of a 
such relationship in order to identify and
understand the sources of possible errors involved in the conversion 
process. 

As a paradigm of a device having a nonlinear internal dynamics, 
we consider the algorithmic analog/digital (A/D) 
converters \cite{converter}. 
The internal dynamics of such converters is described by well-known 
one-dimensional chaotic maps \cite{Ke95}, 
the Bernoulli shift
\begin{equation}
x_{n+1}=G_1(x_n)=2x_n\quad|_{{\rm{mod}}~1}
\label{Bern}
\end{equation}
and the tent map
\begin{equation}
x_{n+1}=G_2(x_n)=1-2|x_n-\frac12|.
\label{tent}
\end{equation}

\noindent
Both maps are characterized by a
positive Lyapunov exponent $\lambda=\ln 2$ and they often serve as model
examples in mathematical work devoted to the theory of
dynamical systems (see e.g. \cite{Gulich}).

The Bernoulli shift is
realized by the circuit shown in Fig. 1(a) \cite{Ke95}.
 This circuit
represents a $K$-bit algorithmic analog/digital (A/D) converter, which is 
a recycling multistage converter with one bit stage.
The converter achieves $K$-bit resolution by
converting the input once and then circulating the residue function
$G_1(.)$ through the stage $(K-1)$ times.
The tent map is realized as a Grey-code algorithmic A/D
converter \cite{Ke95} with a folding circuit
having a transfer characteristic given by Eq. (\ref{tent})
as shown in Fig. 1(b).

To analyze the input-output relationship 
of such analog/digital converters from the
point of view of the theory of dynamical systems, we apply 
concepts from symbolic dynamics. In analyzing the conversion 
process, we use the 
sensitive dependence on the initial conditions, a fundamental property 
of Eqn.(\ref{Bern}) and (\ref{tent}), to argue that 
precise analog/digital conversion is possible.
We identify the sources of errors
in such converters and
present a simple method of error reduction.

Suppose we need to measure the quantity $X$, and let
${\cal X}$ denote the set of all possible values of $X$.
Let us consider an unknown value $x_1$ of $X$, which is the analog 
input of the converter, 
as an initial condition that is iterated internally using the chaotic map
$G$ generating the sequence
$\{x_k\}=x_1 x_2 \dots x_L$ with $x_k=G^{k-1}(x_1)$ and
$k=1,2,\dots,L$.
To describe such an orbit in 
terms of a {\sl symbolic sequence}, 
we need to choose a partition by coarse-graining the space
${\cal X}$. In the case corresponding to the maps
(\ref{Bern}) and (\ref{tent}), we split the space
${\cal X}=[0,1]$ into two cells whose partition is the critical point
$x_c=1/2$.  With any given orbit $\{x_k \}$ one can associate
its symbol sequence $\{a_k\}$ using a two-letter alphabet defined as 
$a_k=0$ for $x_k \le x_c $ and $a_k=1$ for $x_k>x_c$.
The sequence $\{a_k\}$ is called the itinerary of $x_1$.
Let ${\cal O }$ be a set of all possible symbolic sequences.

The process of conversion is pregnant with meaning if
there exists a one-to-one mapping between ${\cal X}$ and ${\cal O}$.
Once the itinerary characterizing the orbit is obtained, we
get a result in form of a digital output
\begin{equation}
y = \gamma(\{a_k\}),
\label{gam1}
\end{equation}
reflecting the unknown  value $x_1$. In general, 
the exact form of the function $\gamma$ depends on the internal dynamics 
of the converter. 
For a dynamics given by the maps
(\ref{Bern}) and (\ref{tent}), it takes a simple form in a binary
 basis 
\begin{equation}
\gamma(\{a_k\}) = 0.c_1c_2...c_L= \sum_{k=1}^{L} c_k2^{-k},
\label{gam2}
\end{equation}
where
$$
c_k=a_k,
$$
for the Bernoulli shift and
$$
c_k=\sum_{i=1}^k a_i\quad|_{{\rm{mod}}~2}
$$
for the tent map \cite{hao89}.
The latter case is sometimes referred to as the
Gray code \cite{Ke95}.

The precision of such an analog/digital converter depends on 
the length $L$ of the
itinerary. In principle, the digital output becomes arbitrarily precise  
for $L$ large
enough, adding one binary digit at each internal iteration of 
Eqs. (\ref{Bern}) or (\ref{tent}), as symbolized in Eq. (\ref{gam2}). 
In practice, the precision of the digital output is limited
by several factors. The following main
sources of errors in the conversion process may be identified:

\begin{description}
\item[(i)] Inaccuracy in modeling: the internal dynamics of the converter 
results in a linear map with slope $s \neq 2$.
\item[(ii)] Partition point placed at $x_c \neq 1/2$.
\end{description}

\noindent
These errors have a significant influence on the precision of the 
digital output.
Errors of the above types destroy the  one-to-one
mapping between ${\cal X}$ and ${\cal O}$ as expressed by Eqs. (\ref{Bern}) 
and (\ref{tent}). In order to deal with errors {\bf{(i)}} and {\bf{(ii)}}, 
we address the following question:
Can we still use converting devices with internal chaotic dynamics 
since errors grow exponentially fast in them?

To visualize the effect of {\it changes in the slope} {\bf{(i)}}, we present in
Fig. 2 the plots of the input--output relations $y=f(x_1)$
for the Bernoulli shift
[Fig. 2(a)] and for the tent map
[Fig. 2(b)]. We choose the discrepancies in the
slope ($\varepsilon_s=2-s$) much larger than in practical applications, to
demonstrate clearly the devil's staircase like character of this
function. Let us denote by $\zeta$ the error of a single conversion,
$\zeta=|y-x_1|$.
The maximal error, $\Delta_s= \max_{x\in
(0,1)}\zeta$, as well as the mean error, $\int_0^1 \zeta (x)dx$,
grow linearly with the discrepancy $\varepsilon_s$.

For our purpose, it is more important the error associated with 
a {\it non-optimal placing of the partition 
point} {\bf{(ii)}} and how it 
influences the precision of the output. Figure 3 presents
the input-output relationship $y(x_1)$ for the Bernoulli map
(a) and the tent map (b) with the correct slope ($s=2$)
and a misplaced partition point $x_c=\frac12 - 2 ^{-M}$
with $M=4$. The data are obtained using
$L=24$ bits at each conversion. The comparison of these pictures
clearly demonstrates the superiority of the tent map over the Bernoulli shift 
for conversion 
purposes. The Bernoulli shift is not
continuous at $x_c=1/2$, which leads to large errors when the 
partition point is displaced. 

To characterize the effect of misplacing the partition point 
quantitatively, we set an arbitrary
tolerance level $\delta<<1$ and compute the cumulative probability
$F(\delta):= \int_{\delta}^1 P(\zeta)d\zeta$, 
the probability that the error exceeds a given  tolerance level.
The dependence $F(\delta)$ obtained for both maps
 is shown in Fig. 4 for a realistically possible 
misplacement $(M=8)$ of the partition point at $x_c=1/2-2^{-8}\approx 0.496$.
Filled circles correspond to the tent map, which is less
vulnerable to imperfections than the Bernoulli shift (empty circles).
%$(\circ)$.

To obtain analytical estimates of the errors, let us assume that 
the dependence shown in Fig. 3 is piecewise linear.
If $x\in D=[x_c,1/2]$ the error occurring for   
the Bernoulli shift is close to $1/2$. For any $x$ belonging to the
first preimage
of this interval, the error is approximately equal to $1/4$. Furthermore,
for $x$ in the $n-$th preimage of the interval $D$,
the error reads $2^{-1-n}$.
The probability distribution of the error
consists thus of the sum of singular Dirac deltas
$P(\zeta)=2^{-M}[\delta(\zeta-1/2)+\delta(\zeta-1/4)+\delta(\zeta-1/8)+\dots]$. 
Calculating the cummulative distribution of error $F(\delta)$, we get
$F(2^{-n})=(n-1)2^{-M}$, $n\ge 1$.
 In the variables used in Fig. 4, this relation reads
$\log_2 F(\log_2 \delta)= -M + \log_2 [-\log_2 \delta -1]$.

A similar reasoning is applicable to the tent map.
For any initial condition $x$ in
the interval of the displacement $D=[x_c,1/2]$, the error
equals $2(1/2-x)$. In this approximation the distribution of error
$P(\zeta)$ consists
of a rectangle $P_0(\zeta)=1/2$ for $\zeta\in[0,2^{-M+1}]$.
There exists two preimages of $D$, for which the
errors belong to $[0,2^{-M}]$.
Their contribution to the distribution of error reads $P_1(\zeta)=1$
for $\zeta\in [0,2^{-M}]$. In an analogous way, the
errors generated by the $n$-th preimages of $D$
are described by $P_n(\zeta)=2^{n-1}$ for $\zeta\in[0,2^{-M+1-n}]$.
To obtain the cumulative error distribution $F(\delta)$ we have to sum
all terms arising from $P_n(\zeta)$, $n=0,1,2,\dots$. Making use of
the sum of a geometric series we arrive at
$F(2^{-M+1-n})= [2n-2+2^{1-n}]2^{-M-1}$. Equivalently, a 
suitable expression for comparison with the numerical data reads
$\log_2 F(\log_2 \delta) =
-M-1 + \log_2 [-2(M+\log_2 \delta) + 2^{M+\log_2 \delta}]$.
Above error estimates, represented  in Fig. 4 by 
continuous and dashed lines, respectively, 
describe the numerical data very well for both maps.

Since the errors caused by the misplacement of the partition point are mostly 
relevant for $x$ at any rational numbers of the
type $k/2^n$, with $n=1,2,\dots$ and $k=1,2,\dots, 2^n-1$, we 
propose a simple {\sl correction scheme}
 based on the idea of a translation of the input value on the
circle (or, equivalently, of the acting map).
Consider a vector ${\bar{v}}=\{v_1, v_2, \dots v_N\}$, in which $v_1=0$ and
all other $N-1$ components are relatively incommensurate numbers
from $(0,1)$. The conversion algorithm with a correction scheme
consists in shifting the unknown quantity and obtaining
$x_1^{(i)}=x_1+v_i |_{{\rm mod}~1}$, converting them by the procedure
described above
and obtaining $N$ different raw digital results $y^{(i)}$,
and performing the reverse shift 
 ${\tilde{y}}^{(i)}=y^{(i)}-v_i |_{{\rm mod}~1}$ with $i=1,\dots, N$.
Some of the results ${\tilde{y}}^{(i)}$, 
may be correct, some can be significantly wrong.
Taking the average of these numbers would be rather meaningless;
we propose to take for $N$ an odd number, order all the results
 ${\tilde{y}}^{(i)}$
and accept the {\sl middle one} as a final result $y$.
To test the practical relevance of this algorithm,  we took ${\bar{v}}=\{0,
\sqrt{2}/2, \sqrt{3}/3, \sqrt{5}/5, \sqrt{7}/7 \}$ and use
 it for both maps
discussed above. As shown in Fig. 4 $N=3$ (triangles) already 
reduces the total error by a factor close to ten. Taking the
middle value of $N=5$ as the resulting output values significantly 
improves even the worse 
converter based on the Bernoulli shift (diamonds). The results for the
continuous tent map with an $N=5$ correction are so good that the cumulative
error, smaller than $2^{-20}$, is not marked in the picture.

The major conclusion of this work is that analog/digital converters with 
chaotic internal dynamics can be used for
precise conversion of physical signals into digital numbers. 
We analyzed the sources of errors
stemming from imperfections of the converter and propose
an error correction algorithm capable of reducing the error of
the conversion. In practice, the final precision of the digital output is 
governed by the accuracy of the auxiliary vector ${\bar{v}}$ but, nevertheless, 
an improvement gained by this procedure can be expected. Besides the errors 
discussed in this paper there are other possible errors leading to 
imprecisions of the digital output. Firstly, the internal dynamics of the 
device is nonlinear instead of piecewise linear. Secondly, the partition 
point location is fuzzy within a finite width $\epsilon >0$. The latter 
problem  has been
recently discussed in the context of communication with chaos
\cite{deltagap}. 

We  summarize the process of analog/digital conversion with internal 
nonlinear dynamics as given by the following steps:

\begin{enumerate}
\item Input of the analog signal $x_1$ to be measured into a nonlinear system
$x_{n+1}=G(x_n)$ as an initial condition.
\item Generation of symbolic sequence $\{a_k\}$ by iterating the 
nonlinear map G(.).
\item  Calculation of  $\gamma(\{a_k\})$ according to
formula (\ref{gam2}).
\item  Obtaining the digital output value $y$ given by Eq. (\ref{gam1}) 
as the result of the measurement.
\end{enumerate}

\clearpage

We thank M. Ogorza{\l}ek and P.~M.~Kennedy for pointing us 
valuable references. 
T. K., K. \.Z. and U. F. thank the University of Maryland for their 
hospitality. 
T. K. and K. \.Z. acknowledge the Fulbright Fellowships
while U. F. acknowledges the support from the Deutsche
Forschungs\-gemein\-schaft and from the Sonderforschungsbereich 555. 
C. G. was supported by DOE and by a 
joint NSF/CNPq grant.

 { }

\newpage

\centerline{\large \bf CAPTIONS}

\vspace{20mm}

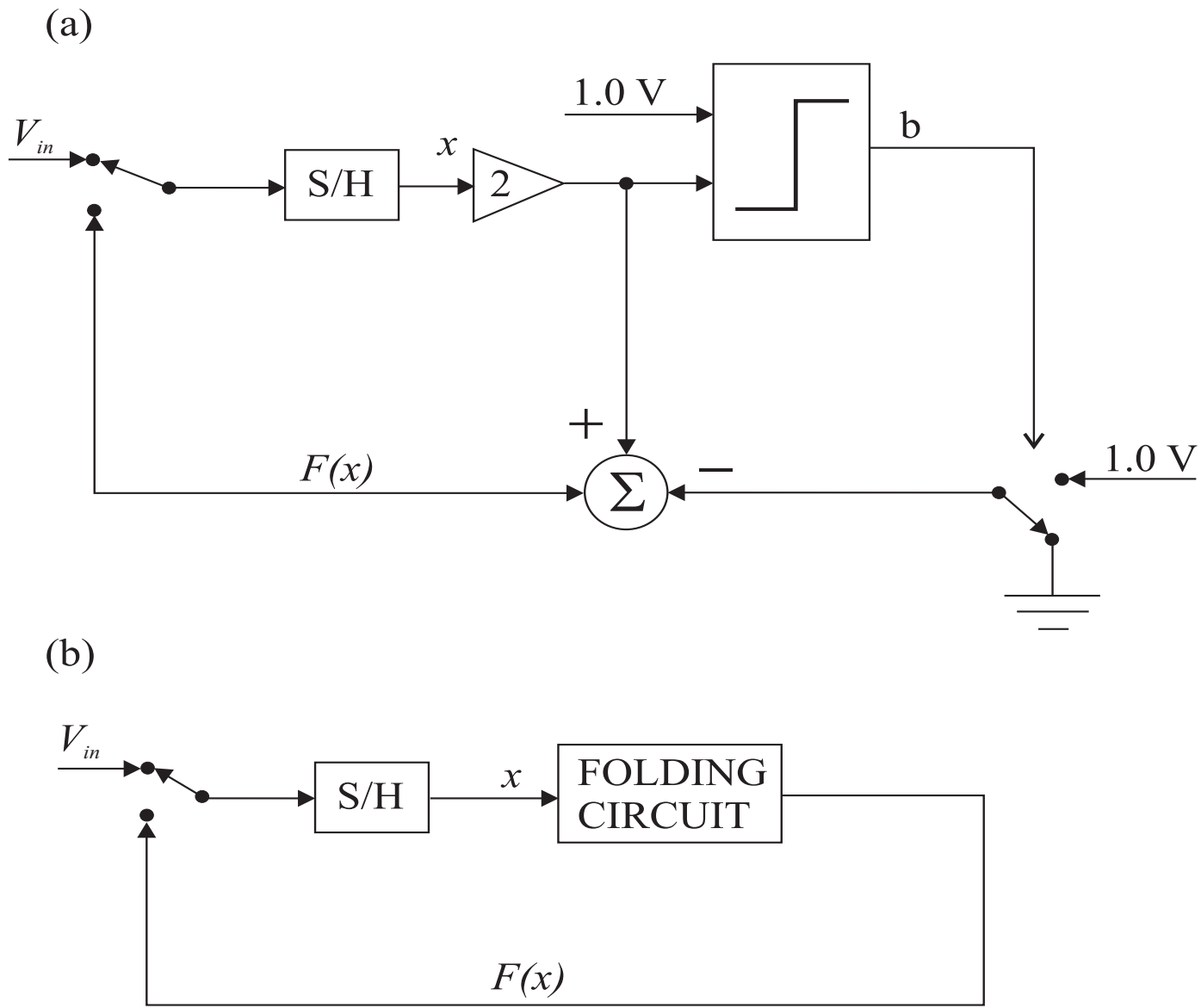
{\bf Figure 1}: The circuit implementation of the Bernoulli shift (a) and the tent map (b).

\vspace{1cm}

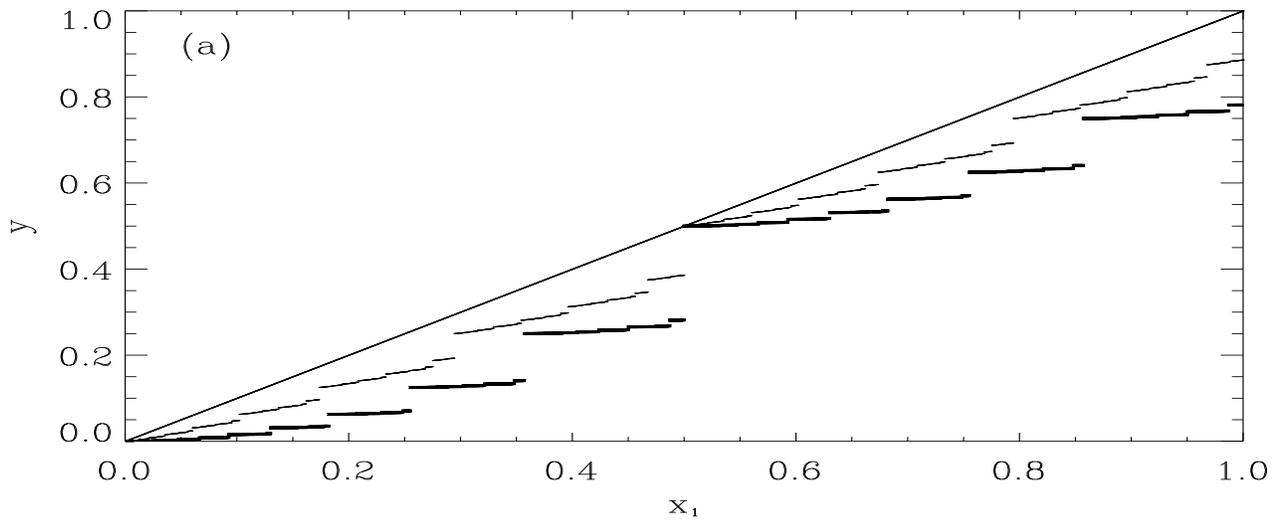
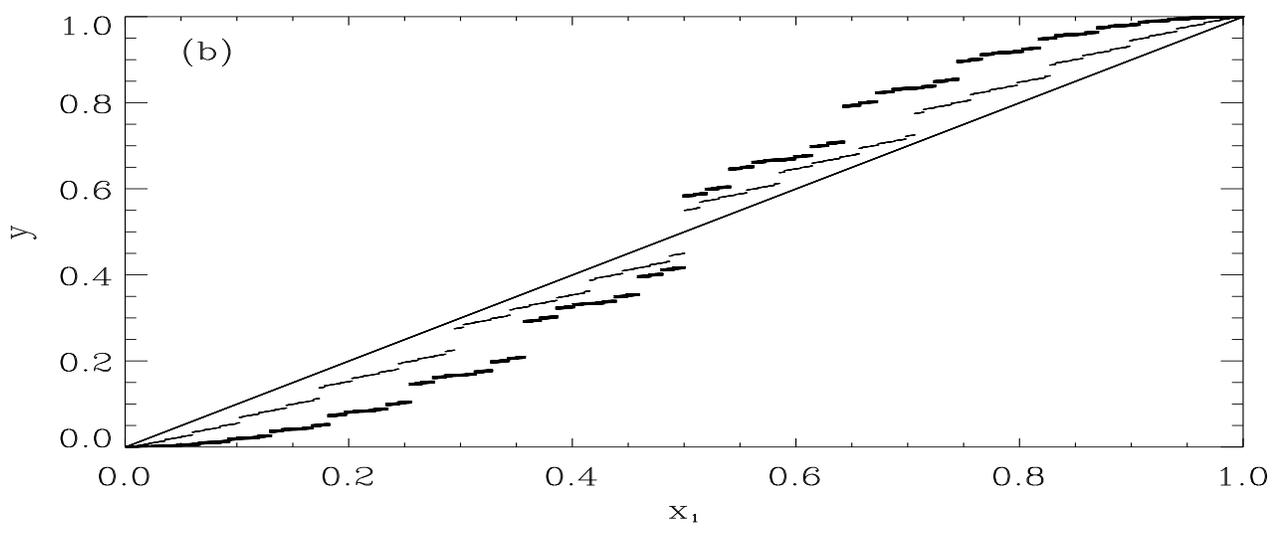
{\bf Figure 2}: Digital output $y$ as a function of the measured input value 
                $x_1$. The conversion scheme is based on the dynamics of the 
                Bernoulli shift (a) and the tent map (b) with the slopes 
                $s=2.0$ (ideal case, diagonal) and non ideal cases 
                $s=1.7$ and $s=1.4$ (thick line).

\vspace{1cm}

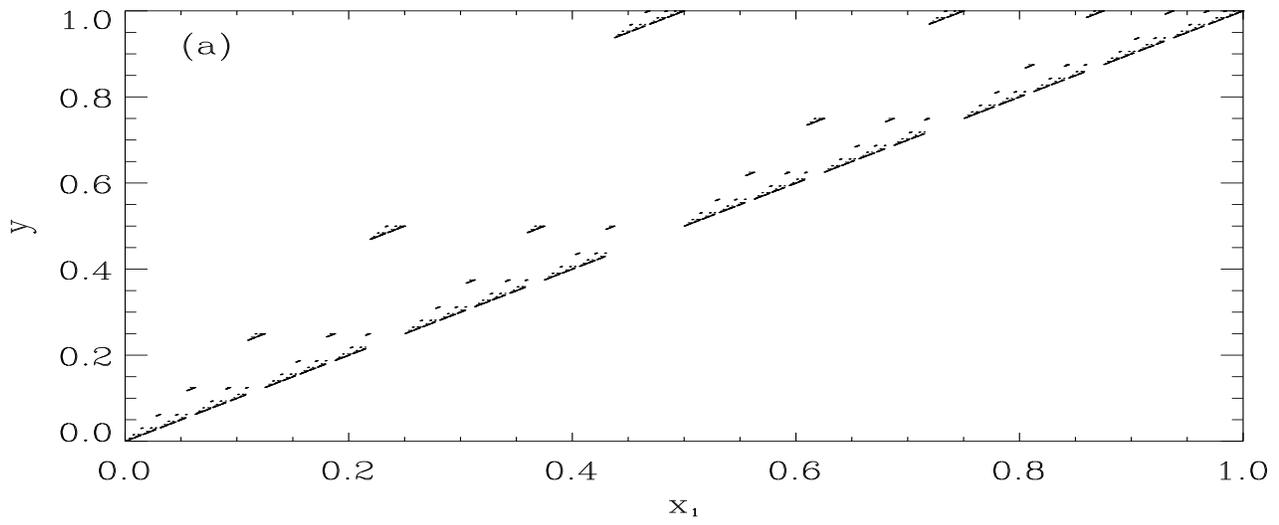
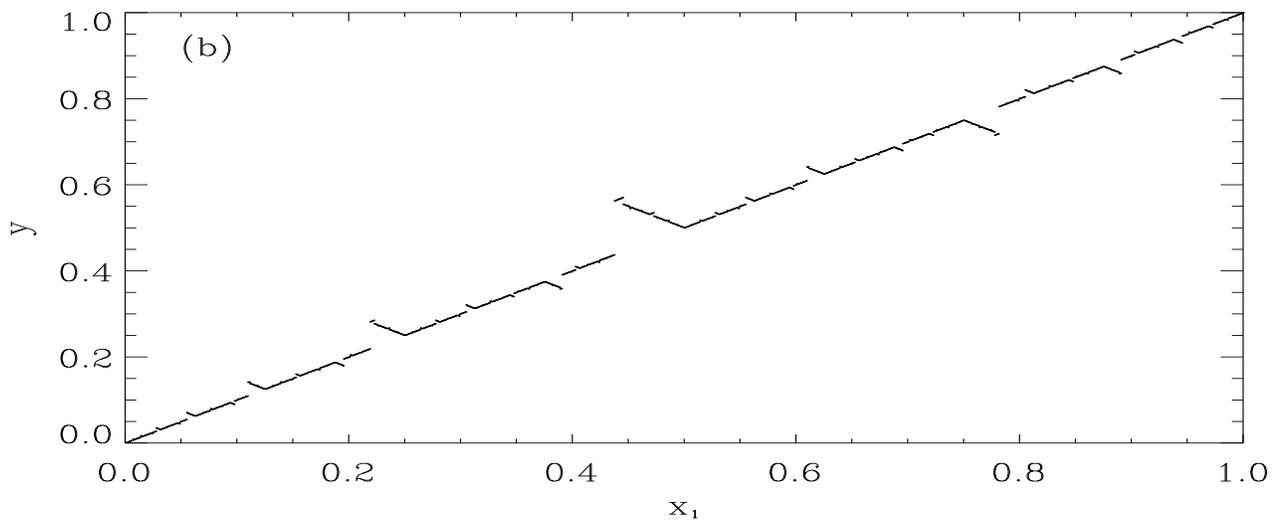
{\bf Figure 3}: As in Fig. 2, the conversions have been performed with systems
                based on the Bernouli shift (a) and the tent map (b).
                Systematic errors are caused by the misplaced partition line
                localized at $x_c=\frac12-2^{-4}$.

\vspace{1cm}

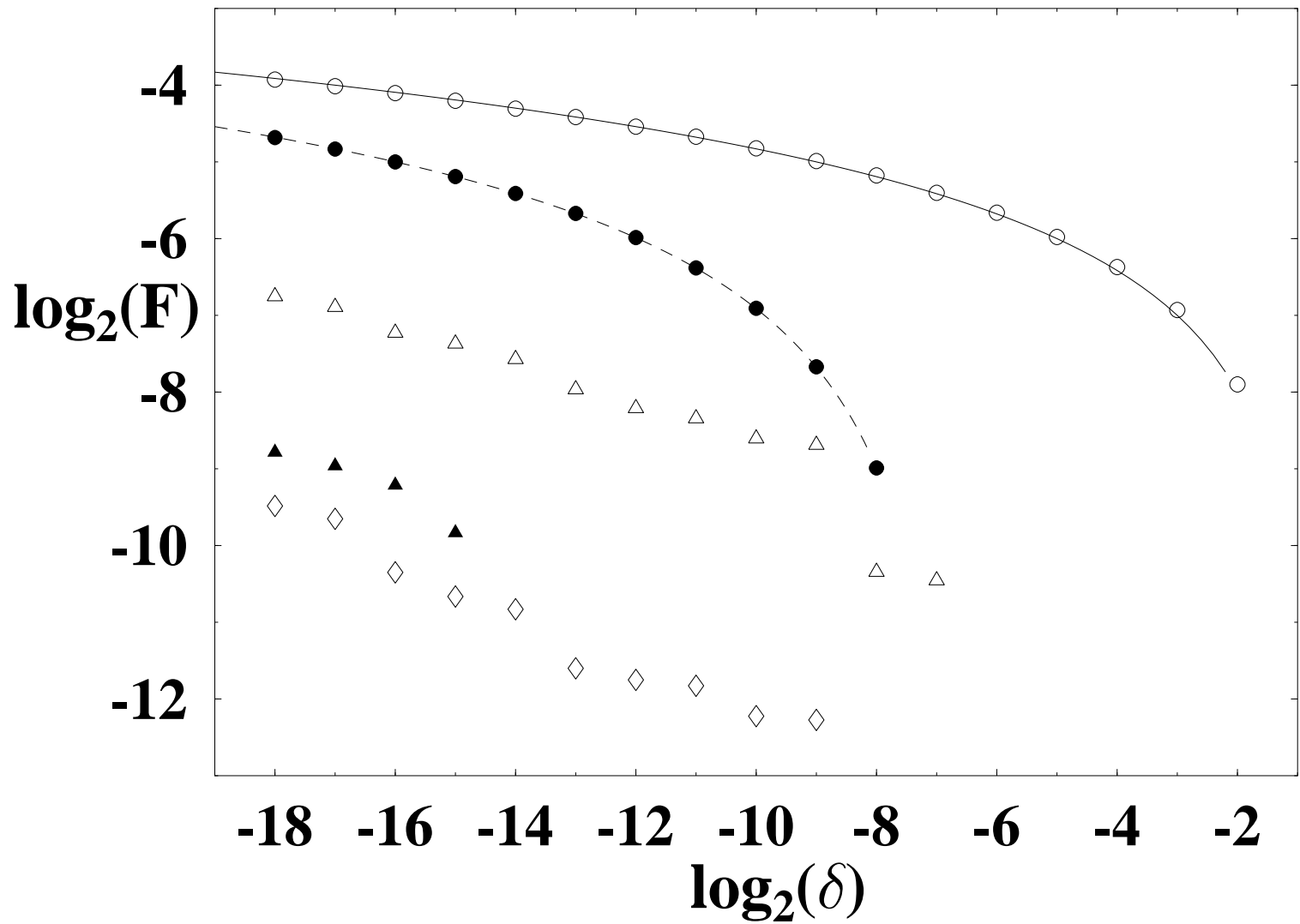
{\bf Figure 4}: Cumulative probability $F$ of error larger than the tolerance
                level $\delta$ obtained for the Bernoulli shift (open symbols) 
                and the tent map (full symbols) with the 
                misplacement of the partition line characterized by $M=8$.
                The circles represent results received by measuring $L=24$
                bits. Triangles and diamonds represent data obtained with 
                $L=24$ by applying the $N=3$ and $N=5$ error--correction 
                scheme.


\begin{thebibliography}{ }

\bibitem{converter}
U. Fiedler and D. Seitzer, IEEE J. Solid-State Circuits {\bf SC-14}, 
547 (1979);
P. W. Li, M. J. Chin, P. R. Gray and R. Castello, IEEE J. Solid-State 
Circuits {\bf SC-19}, 828 (1984);
C. S. G. Conroy, {\sl A High-Speed Parallel Pipeline A/D Converter Technique 
in CMOS}, PhD-thesis, University of California at Berkeley;
P. E. Pace, P. A.Ramamoorthy and D. Styer, IEEE Trans. Circuits Syst.-II 
{\bf 41}, 373 (1994);
G. W. Roberts, {\sl Circuits and Systems Tutorials}, IEEE ISCAS'94, 
London, 534 (1994).

\bibitem{Ke95} M. P. Kennedy, Int. J. Bifurc. \& Chaos, {\bf 5},
891  (1995).

\bibitem{Gulich} D. Gulick, {\sl Encounters with Chaos},
  McGrawHill, New York 1992.

\bibitem{hao89} H. Bai-Lin, {\it Elementary Symbolic Dynamics} (World
Scientific: Singapore, 1989)

\bibitem{deltagap}
E. Bollt, Y.-C. Lai, and C. Grebogi, Phys. Rev. Lett. {\bf 79}, 3787 (1997);
J. Jacobs, E. Ott and B. R. Hunt, Phys. Rev. E {\bf 57}, 6577 (1998);
E. Bollt, Y.-C. Lai, Phys. Rev. E {\bf 58}, 1724 (1998);
K. {\.Z}yczkowski and E. Bollt, {\sl preprint chao-dyn/9807013}.


 \end{thebibliography}
\end{document}